\newcommand{\doublequotes}[1]{``#1''}

\documentclass[a4paper,fleqn]{cas-dc}

\usepackage[authoryear,longnamesfirst]{natbib}

\usepackage{hyphenat}
\usepackage{float}

\begin{document}
\let\WriteBookmarks\relax
\def\floatpagepagefraction{1}
\def\textpagefraction{.001}

\shorttitle{KG enhanced RAG for FMEA}

\shortauthors{Lukas Bahr et~al.}

\title[mode = title]{Knowledge graph enhanced retrieval-augmented generation for failure mode and effects analysis}


%
\author[1,3]{Lukas Bahr}[orcid=0009-0009-5391-0685]

\author[2]{Christoph Wehner}

\author[3]{Judith Wewerka}
\author[3]{José Bittencourt}
\author[2]{Ute Schmid}
\author[4,5]{Rüdiger Daub}

\credit{Conceptualization of this study, Methodology, Software}
\affiliation[1]{organization={TUM School of Engineering and Design, Technical University of Munich},
    country={Germany}}
\affiliation[2]{organization={Cognitive Systems Group, University of Bamberg},
    country={Germany}}
\affiliation[3]{organization={Digitization Department for Battery Production, BMW Group},
    country={Germany}}
\affiliation[4]{organization={Institute for Machine Tools and Industrial Management, Technical University of Munich},
    country={Germany}}
\affiliation[5]{organization={Fraunhofer Institute for Casting Composite and Processing Technology IGCV, Fraunhofer},
    country={Germany}}


\begin{abstract}
Failure mode and effects analysis (FMEA) is an essential tool for mitigating potential failures, particularly during the ramp-up phases of new products. However, its effectiveness is often limited by the reasoning capabilities of the FMEA tools, which are usually tabular structured.
Meanwhile, large language models (LLMs) offer novel prospects for advanced natural language processing tasks. However, LLMs face challenges in tasks that require factual knowledge, a gap that retrieval-augmented generation (RAG) approaches aim to fill. RAG retrieves information from a non-parametric data store and uses a language model to generate responses.
Building on this concept, we propose to enhance the non-parametric data store with a knowledge graph (KG).
By integrating a KG into the RAG framework, we aim to leverage analytical and semantic question-answering capabilities for FMEA data.
This paper contributes by presenting set-theoretic standardization and a schema for FMEA data, an algorithm for creating vector embeddings from the FMEA-KG, and a KG-enhanced RAG framework.
Our approach is validated through a user experience design study, and we measure the precision and performance of the context retrieval recall.
\end{abstract}


\maketitle

\section{Introduction}
\label{sec:introduction}
A global and close-to-simultaneous start of production for a new product challenges multiple and interdisciplinary teams along the vertical and horizontal value chain. Quality assurance teams are particularly concerned during the start of production with poorly controlled processes \cite{colledani_production_2018}.
Furthermore, processes are often complex, and cause-and-effect relationships are difficult to identify, for example, due to various operations, coordination, or communication barriers in a global production network \cite{possner_anwendung_2024, wehner_interactive_2023}. For this reason, experience gained in pre-series with regard to possible problems during the ramp-up and operation of production systems must be transferred to the series plants in a global production network using user-centered approaches \cite{eirich_manuknowvis_2023}. 

Failure mode and effects analysis (FMEA) is one tool to avoid potential failures during a ramp-up phase. FMEA is a risk analysis tool that focuses on systematically identifying failures and preventing defects, for example, in the process chain or the product's design \cite{isoiec_risk_2019, iec_iec_2018}.
Typically, many stakeholders are involved in complex products, leading to different and incoherent FMEA approaches that make it difficult to reason over the FMEA analysis between the different development units \cite{ozarin_bridging_2013}. Furthermore, FMEA documents become quickly inextricable as actors maintain inconsistent data entries, leading to documentation that is difficult to evaluate with unclear completeness and integrity \cite{carlson_effective_2012}. Lastly, FMEA results are usually tabular structured and lack reasoning capabilities, such as finding generalizations in error patterns, summarizing, or analytic guidance for error prevention.

Recent advances in large language models (LLMs) enable fine-tuning of custom data sets. This can be used for information retrieval and interactions with facts of the FMEA. 
LLMs are commonly trained on large amounts of text and show great potential, among others, in tasks such as language comprehension or answering questions \cite{brown_language_2020}.
However, LLMs often fail when factual knowledge is required \cite{pan_unifying_2024}. 
Designing LLMs for tasks that require factual knowledge remains an open research question. In \cite{shuster_retrieval_2021}, the authors show that retrieval-augmented generation (RAG) outperforms language models for question-answering (QA) tasks that require factual precision. The RAG architecture stores the information in a non-parametric store and utilizes a language model to generate the answer. 
A non-parametric store is a flexible and schema-less database solution for storing and retrieving data.
However, current RAG approaches face difficulties extracting analytical data due to the loss of symbolic meaning attached to numerical values.
This limits drawing conclusions from the numerical information of the FMEA.
For example,\textit{\doublequotes{What failure cause has the highest risk priority number for the process step X?}} cannot be correctly inferred from the non-parametric data store.

One possible solution is to embed the non-parametric data stored in a knowledge graph (KG). KGs store symbolic facts by representing data as nodes and relations in a directed, labeled multi-graph \cite{hogan_knowledge_2021}. The symbolic structure of the KG contextualizes the data and allows interpretable reasoners to derive novel insights from the data \cite{schramm_comprehensible_2023}. 
Prominent tools such as search engines (Google, Bing, etc.) or query answering services (Apple Siri, Amazon Alexa, etc.) have recognized the benefits of semantic technologies \cite{noy_industry-scale_2019, nigam_review_2020}. 

\begin{figure}[t]
\centering
\includegraphics[width=0.5\textwidth]{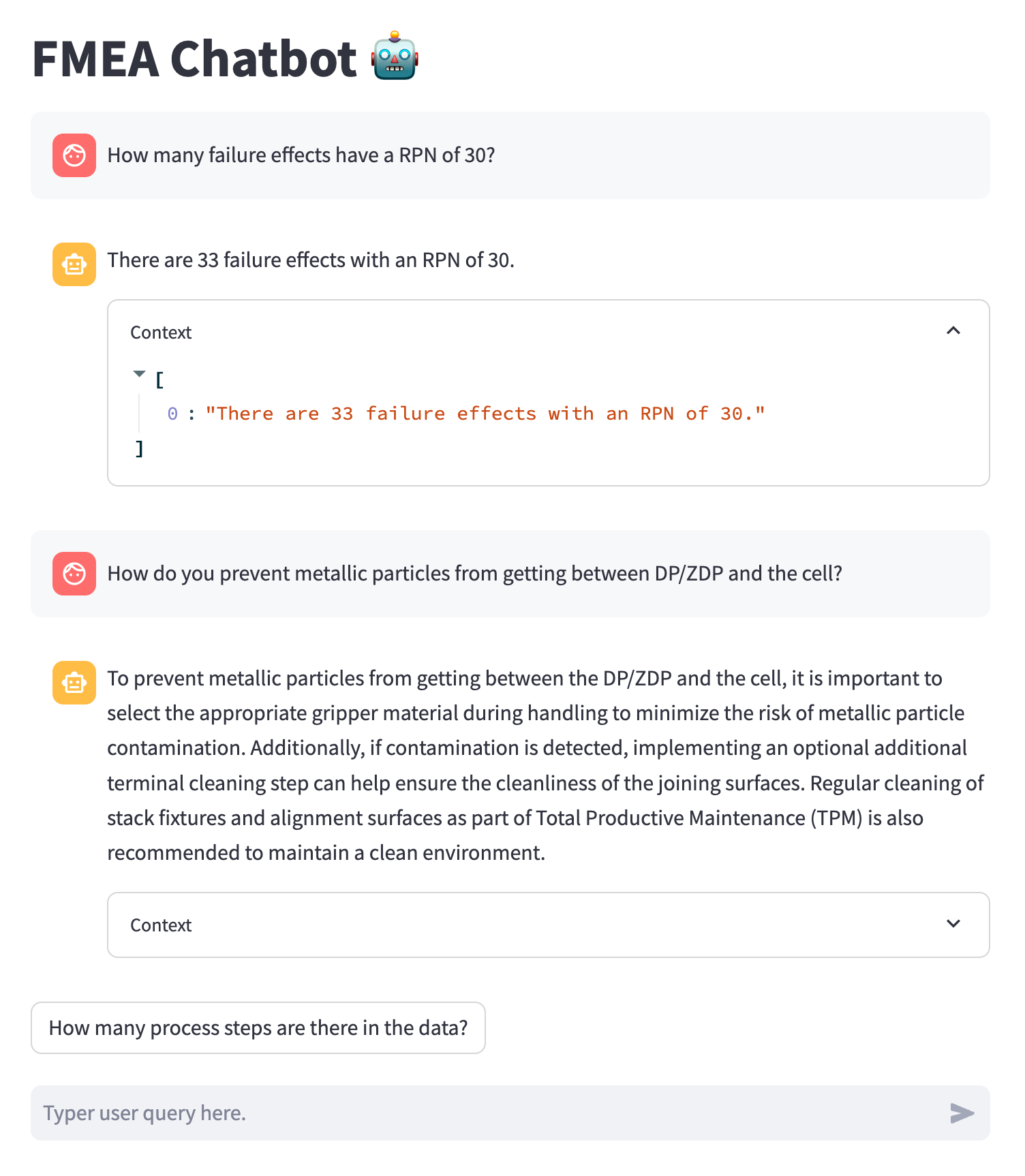}
\caption[FMEA chatbot]{Screen capture of the FMEA chatbot. The user interface is kept simple. Clicking on context reveals the information retrieved from the FMEA-KG.}
\label{fig:fmea_chatbot}
\end{figure}

This paper contributes by integrating KGs within a RAG framework for user-friendly and analytic QA on FMEA data. 
This paper formalizes a set-theoretic standardization of the FMEA, enabling any FMEA to be mapped into a KG, regardless of the tools used for the FMEA's creation or storage. To transpose this set-theoretic standardization into a KG, the paper proposes a schema for any FMEA.
Subsequently, the paper suggests an algorithm for traversing the FMEA-KG to create vector embeddings. Finally, the paper demonstrates how to integrate the non-parametric data store into a KG, resulting in a KG-enhanced RAG (KG-RAG) framework.
The KG-RAG framework enables retrieving human-understandable information while allowing analytical capabilities through the KG graph query. 
A user experience design study provides information on the effectiveness of the approach. We further measure the performance of context recall and precision.

\section{Related work} \label{sec:related_work}
The following briefly introduces the influential work on FMEA, KG, LLM, and RAG relevant to this paper.
The methods defined in ISO/IEC 31010 \cite{isoiec_risk_2019} dominate the risk assessment in manufacturing, inter alia: the five whys method \cite{serrat_five_2017}, fault tree analysis \cite{ruijters_fault_2015}, and failure mode and effect analysis (FMEA) \cite{iec_iec_2018, iec_iec_2006}. 
FMEA is a systematic methodology to analyze potential failure modes within a system to classify the severity and likelihood of failures \cite{schmitt_qualitatsmanagement_2015}. This structured approach aims to identify, prioritize, and mitigate risks associated with system failures \cite{stamatis_failure_2003}.

However, challenges arise when dealing with convoluted systems, such as in manufacturing \cite{henshall_systems_2014}. In particular, incoherent FMEA approaches remain due to the intersection between different domains \cite{ozarin_bridging_2013}, difficulties in identifying and managing errors \cite{hunt_failure_1995}, and large and poorly structured FMEA documents \cite{carlson_effective_2012}. Some methods aim to facilitate the accessibility of FMEA documents, such as integrating FMEA with quality function deployment \cite{almannai_decision_2008}. Recent research has integrated game theory into FMEA to improve prioritization methods during its establishment. The approach in \cite{yazdi_enhancing_2024} employs a zero-sum game strategy to rank failure modes and dynamically refine risk assessments. The best-worst method utilizes Pythagorean fuzzy uncertain linguistic variables and Nash equilibrium principles \cite{li_nash_2024}.

Instead of storing the FMEA data in a classical relational (or non-relational) database, KGs offer versatile access to information.
KGs are a way to store and retrieve facts \cite{hogan_knowledge_2021}, making them a crucial technology to answer questions based on factual knowledge \cite{shuster_retrieval_2021}. Novel KG tools allow for the storage of vector embeddings \cite{mittal_thinking_2017}. In \cite{schneider_course_1973}, KGs are initially conceived as semantic networks and have diverse applications, including ventures into human language through projects such as WordNet \cite{miller_wordnet_1995}.
Subsequently, a multitude of private companies and academic institutions have embraced KGs for a wide array of applications \cite{hogan_knowledge_2021}, such as DBpedia \cite{auer_dbpedia_2007}, Freebase \cite{bollacker_freebase_2008}, or Google KG \cite{singhal_introducing_2012}. 
The default method of interacting with formalized knowledge in KGs is a structured query language (e.g., Cypher, SPARQL) \cite{neo4j_neo4j_nodate, hogan_knowledge_2021}, which requires domain knowledge of the KG schema for effective data extraction, making interactions cumbersome for non-domain experts.

Advances in LLMs allow for fine-tuning custom data sets, enabling a retrieval approach for FMEA observations. The LLM's transformer architecture enables training with its self-attention mechanism for multiple sequences that simplify, for example, the optimization process and minimize the risk of vanishing gradients \cite{lin_survey_2022, brown_language_2020}. LLMs are trained on large amounts of data and tackle various tasks in the field of natural language processing (NLP), such as language comprehension or QA, and have also found adoption in the manufacturing domain \cite{brown_language_2020, holland_large_2024}.
These models have shown significant advantages over non-transformer architectures, such as recurrent neural networks (RNNs) and long short-term memory (LSTM) networks, in terms of scalability, improved contextual embeddings, and overall performance \cite{lin_survey_2022}.
OpenAI's GPT (Generative pre-trained transformer) models, such as GPT-4 \cite{openai_gpt-4_2023}, are noteworthy for their application of large language models (LLMs) in dialogues, providing human-like conversational capabilities \cite{zhao_survey_2023}. 
However, LLMs struggle in tasks requiring factual knowledge, making LLMs unreliable in expert domains critical to manufacturing \cite{pan_unifying_2024}. Furthermore, LLMs tend to hallucinate, leading to misinformation and making them unreliable in tasks where accurate knowledge recall is crucial \cite{ji_survey_2023, rawte_troubling_2023}. Additionally, LLMs are a black-box architecture that results in a lack of interpretability and complicates understanding the reasoning of the model \cite{danilevsky_survey_2020, wehner_explainable_2022}.

Rather than fine-tuning a black-box model on FMEA data, LLMs allow retrieving vector embedding of its input, permitting architectures for better explainable methods such as RAG \cite{lewis_retrieval-augmented_2020}. 
RAG approaches save information chunks as vector embeddings in a non-parametric store and utilize natural language generation methods to generate the answer from the store \cite{li_survey_2022}. This approach allows for the composition of responses that minimize hallucinations, making the responses of chat-based LLMs more fact-based.
However, since the information is represented as a vector, the literal and symbolic meaning of the numerical values is lost, which remains a challenge for the current RAG approaches \cite{gao_retrieval-augmented_2023}. For example, retrieving the highest-risk priority number of the FMEA likely outputs the wrong information, as the symbolic meaning of the number is embedded in a vector.

Vector embeddings represent the strings in a high-dimensional vector space.
In \cite{mikolov_efficient_2013}, the authors initially propose embedding words into a meaningful representation by taking advantage of the vector dimensionalities.
Novel advanced methods, such as BERT \cite{devlin_bert_2019}, adopt transformer-based self-supervised language models, resulting in richer representations of the language structure. In \cite{neelakantan_text_2022, openai_gpt-4_2023}, it is shown that pre-training on a sufficiently large batch size can lead to high-quality vector representations.
\section{QA for FMEA} \label{sec:methodology}
In this paper, we introduce the KG-enhanced RAG (KG-RAG) framework for analytical and semantic QA on FMEA data. Unlike fine-tuning an LLM, the KG-RAG framework allows dynamic data updating without relearning and basic numerical analytics, like identifying the failure mode with the highest risk priority number.
In Section~\ref{subsec:standardizing_fmea}, we formalize FMEA set-theoretically, enabling transposition into the FMEA knowledge graph described in Section~\ref{subsec:embedding_fmea}. We also outline an algorithm for computing vector embeddings from the FMEA-KG. In Section~\ref{subsec:rag}, we introduce the KG-RAG framework, which retrieves the FMEA context through vector search, enhanced with KG graph queries, for customized user query results.

\subsection{Set-theoretic formalization of FMEA} 
\label{subsec:standardizing_fmea}
In the context of manufacturing systems, any FMEA can be formally defined using a set-theoretic framework. An FMEA consists of a set $F$ where each element represents a distinct failure mode \cite{automotive_quality_and_process_improvement_committee_potential_2021}. A failure mode is the specific manner in which a process, product, or system could potentially fail to perform its intended function. Formally, let $F =\{f_1,f_2,...,f_n\}$ be the set of all identified failure modes. Each failure mode $f \in F$ is associated with the following elements \cite{automotive_quality_and_process_improvement_committee_potential_2021}:
\begin{itemize}
    \item \textbf{Failure effect} $E(f)$: The effect or impact of the failure mode on the system.
    \item \textbf{Failure cause} $C(f)$: The underlying cause or reason for the failure mode.
    \item \textbf{Failure measure} $M(C(f))$: The measure used to evaluate the failure mode.
\end{itemize}
Furthermore, there is a mapping $A: F(f) \rightarrow P$ from the failure mode $F(f)$ to its appearance within the manufacturing line, where $P$ denotes the occurrence in the manufacturing process, such as process step or station.
The FMEA employs further mappings that evaluate the severity, occurrence probability, and detection probability with a rating $R = \{1, \dots, 5\}$, for which a higher rating indicates a higher criticality for the failure mode \cite{stamatis_failure_2003}. The risk priority number (RPN) gives the overall risk level rating of the failure mode. We follow the mappings:
\begin{itemize}
    \item \textbf{Severity} $S$: A mapping $S: E(f) \rightarrow R$, which takes a failure effect $E(f)$ and returns the severity rating.
    \item \textbf{Occurrence probability} $O$: A mapping $O: C(f) \rightarrow R$, which takes a failure cause $C(f)$ and returns the probability rating of the event.
    \item \textbf{Detection probability} $D$: A mapping $D: M(C(f)) \rightarrow R$, which takes a failure measure $M(C(f))$ and returns the detection rating probability.
    \item \textbf{Risk priority number} $RPN$: A mapping $RPN: (S, O, D) \rightarrow U$, which takes $S$, $O$, and $D$ and returns the overall risk level rating $U = S \cdot O \cdot D$ \cite{iec_iec_2018, iec_iec_2006}.
\end{itemize}

Defining the core elements and mappings of an FMEA in this manner allows any specific FMEA to be integrated into the set-theoretic formalization. This approach ensures consistency and facilitates systematic analysis across different FMEAs. In addition, it enables the transposition of any FMEA into a KG for querying and vector search.

\subsection{Modeling FMEA data within a KG}
\label{subsec:embedding_fmea}
We propose a two-step process to store and embed FMEA data in a KG. Typically, FMEA data are structured in a tabular format within a database.
First, the tabular data structure is mapped to the previously proposed set-theoretic FMEA formalization. 
The standardized set helps to transpose the FMEA data into a KG structure for which a schema is defined.
Second, the FMEA data must be aggregated, embedded, and attached to the KG to enable the graph structure for vector search.

\begin{figure*}[t]
\centering
\includegraphics[width=0.8\textwidth]{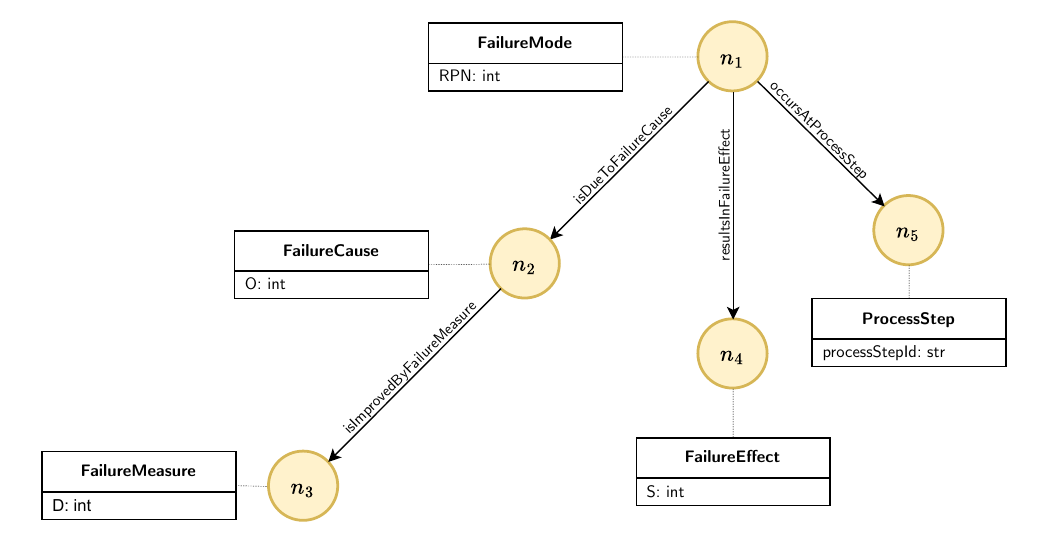}
\caption[Schema FMEA]{Overview of the proposed schema for the FMEA, including the symbols, literals, and relations. The nodes ($n_1 - n_5$) represent a failure mode and illustrate how depth-first search is conducted on the tree-like graph structure.}
\label{fig:fmea_schema}
\end{figure*}

\subsubsection*{Transposing the FMEA data into a KG}
\label{subsubsec:transposing}
Consider a KG $\mathcal{G}$ to be a directed graph defined by
\begin{align}
\mathcal{G} =(N,R,T),
\end{align}
where $n \in N$ is the set of nodes and $r \in R$ are the relations of the graph \cite{schneider_course_1973, hogan_knowledge_2021}. The set $T$ consists of facts (triples) in the form $T = (n_h, r, n_t)$, where $n_h$ is the head node, $r$ is the relation, and $n_t$ is the tail node \cite{hogan_knowledge_2021}.
KGs enhanced with literals $l \in L$ add descriptive features to the graph. These literals provide additional information about the nodes and their relations, transforming $\mathcal{G}$ into a knowledge graph with literals. In this context, a node is represented as a tuple $n = (s_n, l_1, \ldots, l_k)$, where $s_n$ denotes the symbol of the node (such as its name), and $l_1, \ldots, l_k$ are literals providing descriptive information \cite{hogan_knowledge_2021}. Literals can be descriptive strings, numeric values, or any other information about the node. Similarly, a relationship is indicated as $r = (s_r, l_1, \ldots, l_m)$, where $s_r$ is the symbol that identifies the relationship $r$. The variables $k$ and $m$ denote the number of literals associated with the nodes and the relations, respectively. 

KG can be structured using a schema that provides a set of classes and rules specific to the domain. This includes establishing a class hierarchy and defining the types of relations, such as their origin and target categories \cite{hogan_knowledge_2021}. 
To formalize an FMEA schema, we map the set-theoretic formalization of an FMEA to the corresponding node $s_n$ and relation $s_r$ symbols of the schema. For $f \in F$, each failure mode corresponds to a subgraph of the FMEA-KG. 
We define the nodes $F(f)$ as \emph{FailureMode}, $E(f)$ as \emph{FailureEffect}, $C(f)$ as \emph{FailureCause}, $M(C(f))$ as \emph{FailureMeasure} and $P$ as \emph{ProcessStep}.
The schema realizes a hierarchy between $F(f)$ and $C(f)$ by the relationship \emph{isDueToFailureCause}, $C(f)$ and $M(C(f))$ by \emph{isImprovedByFailureMeasure}, $F(f)$ and $E(f)$ by \emph{resultsInFailureEffect}.
The mapping $A: F(f) \rightarrow P$ corresponds to the relation \emph{occursAtProcessStep}, linking the failure effect to a process step.
Other mappings such as $RPN$, $S$, $O$, and $D$ are numerical literals associated with their respective nodes within the KG schema. Figure~\ref{fig:fmea_schema} shows a visualization of the complete FMEA schema.

The modeling of FMEA as KG allows the usage of graph algorithms, such as reasoning, clustering, or pathfinding, and a query language that facilitates the retrieval of complex joint information \cite{hogan_knowledge_2021}. Although there are sophisticated reasoning methods within KGs, e.g., ruled-based or distributed representation-based inference methods, the methods do not scale with the KG's size or have difficulty retrieving deeper compositional information \cite{chen_review_2020}. Particularly poorly formulated queries pose challenges for reasoning methods. To address this issue, we propose to embed information chunks of the FMEA data as vector representations, enabling vector-based reasoning.

\subsubsection*{Creating vector embeddings of the FMEA-KG}
To ensure meaningful information extraction from the vector search, vector embeddings must represent the entire FMEA graph. For this, we build information chunks from the KG that capture the underlying structure of the FMEA.
For example, an inquiry such as\textit{\doublequotes{What is the consequence of X on Y?}} is not explicit and depends on the context \emph{X} and \emph{Y}. The results could differ by a slight change with the context of the inquiry. Hence, adopting an efficient approach that encompasses all potential information within the FMEA data is essential.

For this, we propose traversing the FMEA-KG by exploiting the tree-like structure of the FMEA-KG (cf. Figure~\ref{fig:fmea_schema}) to transform the hierarchical structure into a sequential structure.
For each \emph{FailureMode} node, we perform a depth-first search (DFS) to obtain the connecting nodes.
Next, we compile a singular string representation of the nodes traversed from the root node. With the generated text chunk, we compute the vector embeddings utilizing any word-to-vector encoder. Algorithm~\ref{algo:fmea_chunk} fully details the steps for getting the vector embeddings. Subsequently, we create a new node $n$ labeled \emph{VectorEmbedding} that indexes and stores the vector embeddings as literal and links it to the corresponding \emph{FailureMode} node.

\begin{algorithm*}[t]
\caption{Get vector embeddings from FMEA graph $\mathcal{G}$}\label{algo:fmea_chunk}
\begin{algorithmic}[1]
\Procedure{getVectorEmbeddings}{$\mathcal{G}$}
\State $\triangleright \text{ } vectorEmbeddings: \text{set of vector embeddings}$
\For{$node \in FailurMode$}
    \State $connectedNodes\gets DFS(node)$
    \State $\triangleright \text{ } chunk: \text{string holding properties of connectedNodes}$
    \For{$properties \in connectedNodes$}
        \State $chunk\gets add(properties)$
    \EndFor
    \State $vectorEmbeddings \gets computeVectorEmbedding(chunk)$
\EndFor
\State \Return $vectorEmbeddings$
\EndProcedure
\end{algorithmic}
\end{algorithm*}

\subsection{KG-RAG framework} \label{subsec:rag}
The LLM leverages its extensive training on vast amounts of data to comprehend language nuances and provide contextually accurate output. We define the operation of the LLM in the following way. LLMs process text input $x$ based on a given context or instruction $c$ to generate an output. $h_{process}$ takes $x$ and $c$ as input and generates a combined input. This combined input is fed into the LLM, which computes the inference by $LLM(h_{process}(x, c))$.

Let $\mathcal{G}$ denote the FMEA-KG proposed in Section~\ref{subsec:embedding_fmea}. Subsequently, we introduce the KG-enhanced RAG framework for information retrieval from $\mathcal{G}$. The framework has two components for the retrieval of information: (i) the graph query language of $\mathcal{G}$ and (ii) vector search.

The KG query language enables retrieving specific information for which the object is known or for basic analytics. Let $w$ denote a user inquiry. For example,\textit{\doublequotes{What is the cumulative risk priority number for the process step X?}} An LLM automatically generates a specific query $q$ adhering to the querying language definition. For this, the LLM generates the graph query using the appropriate graph query language, such as SPARQL, by leveraging the context $c$ provided by the graph's semantics and instruction. The graph query $q = LLM(h_{process}(w, c))$ is then executed on $\mathcal{G}$. The process can be expressed by $\mathcal{Q}: q \rightarrow \mathcal{R_{\text{query}}}$, where $\mathcal{R_{\text{query}}}$ is the set of retrieved query results from $\mathcal{G}$.

For queries that do not return results due to unknown semantic information, such as wrong node label conventions in $w$ or failures in the automatic query generation, we provide a vector search based on vector embeddings in $\mathcal{G}$. 
Let $v_w$ denote the vector representation of $w$, and $\{\mathbf{v}_i\}_{i=1}^{\mathcal{I}}$ be the set of vector embeddings in the database, where $\mathcal{I}$ is the total number of embeddings in $\mathcal{G}$.
We define a similarity function $\mathcal{S}: (\mathbf{v}_w, \{\mathbf{v}_i\}_{i=1}^{\mathcal{I}}) \rightarrow \{\mathbf{v}_{i_1}, \mathbf{v}_{i_2}, \dots, \mathbf{v}_{i_k}\}$, which returns the vectors $k$ most similar to $\mathbf{v}_w$. The results of the most similar vectors are given by $\mathcal{R_{\text{vector}}}$.
The similarity function is given by the cosine function $S: (\mathbf{v}_w, \mathbf{v}_i) \rightarrow \cos(\theta) = (\mathbf{v}_w \cdot \mathbf{v}_i) / (\|\mathbf{v}_w\| \|\mathbf{v}_i\|)$, which measures the similarity between $\mathbf{v}_w$ and $\mathbf{v}_i$ \cite{camacho-collados_word_2018}. 
It ranges from $-1$ to $1$, with $-1$ indicating completely opposite vectors and $1$ indicating similar vectors. 

\begin{figure*}[t]
\centering
\includegraphics[width=0.8\textwidth]{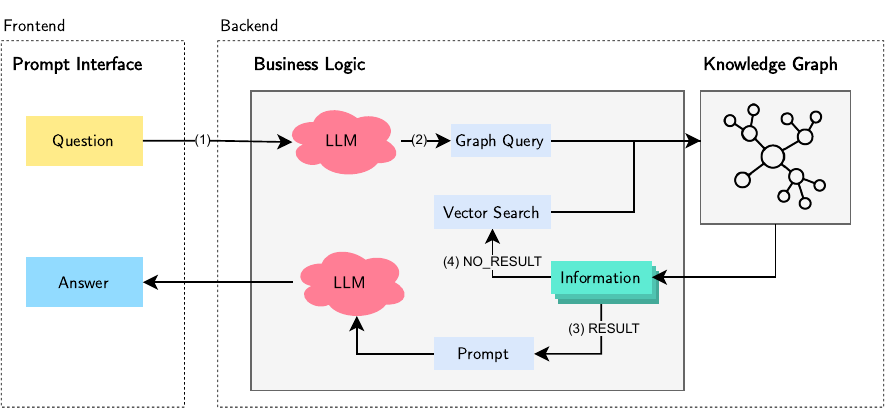}
\caption[Architecture of the RAG framework]{Sketch of the proposed KG-RAG framework. The information is retrieved either by utilizing the graph query language of the KG or by using vector search. The framework employs an LLM to generate the query and to serve the result.}
\label{fig:architecture_overview}
\end{figure*}

The general KG-RAG framework operates with two pipelines: one for retrieving information from the FMEA-KG given the user inquiry $w$ and another for serving the information to the user. Figure~\ref{fig:architecture_overview} summarizes the framework. Initially, the user inquiry $w$ is fed into an LLM (cf. Fig.~\ref{fig:architecture_overview} (1)). The LLM generates a graph query $q$, considering $w$ and the graph's semantic structure of the FMEA schema (cf. Fig.~\ref{fig:architecture_overview} (2)). 
If $\mathcal{Q}(q)$ yields a result, the information retrieval is complete (cf. Fig.~\ref{fig:architecture_overview} (3)). If $\mathcal{Q}(q)$ does not output any results, we perform a vector search $\mathcal{S}(v_q, \{v_i\}_{i=1}^{\mathcal{I}})$ to retrieve the $k$ most similar vectors (cf. Fig.~\ref{fig:architecture_overview} (4)).
The results from the retrieval process, either $\mathcal{R_{\text{query}}}$ or $\mathcal{R_{\text{vector}}}$ are input to the LLM to generate a response addressing the user's inquiry embedded in the context $c$. This can be expressed by $LLM(h_{process}(\mathcal{R}, c)).$

\subsubsection*{Handling of epistemic uncertainty}
LLMs are prone to hallucination when faced with epistemic uncertainty when they lack sufficient training data for similar examples \cite{hullermeier_aleatoric_2021}. By default, LLMs cannot indicate uncertainty and generate plausible-sounding responses regardless of confidence. The proposed KG-RAG framework approach addresses this by relying on retrieved context rather than model parameters for answer generation. 
It enables users to assess uncertainty by displaying retrieved contexts. 
Graph query-based contexts are highly reliable, as they derive from symbolic manipulation and factual data in the FMEA. 
For vector-based search, the KG-RAG approach provides the cosine distance alongside the retrieved context, signaling semantic proximity to the user query. Greater distances indicate higher epistemic uncertainty.
\section{Experiments}
\label{sec:experiment}
We evaluate the KG-RAG framework for FMEA QA, focusing on the capability of context retrieval. To facilitate user interaction with the KG-RAG framework, we introduce a simple FMEA chatbot interface. Section~\ref{subsec:implementation_details} covers our implementation details. Sections~\ref{subsec:methodology} and \ref{subsec:results} detail the methodology and results of the user experience design study. Section~\ref{subsec:online_evaluation} assesses the chatbot's numerical data retrieval effectiveness by measuring context precision and recall. Finally, Section~\ref{subsec:discussion} discusses the results.

\subsection{Implementation details} \label{subsec:implementation_details}

\begin{table*}[t]
    \centering
    \caption[FMEA example data]{Example data from FMEA used in our experiments. The FMEA contains data to mitigate risks in the production of HVS.}
    \label{tab:fmea_example_data}
    \begin{tabular*}{\textwidth}{@{\extracolsep{\fill}}
    p{0.12\textwidth} p{0.12\textwidth} p{0.12\textwidth} p{0.03\textwidth} p{0.12\textwidth} p{0.03\textwidth} p{0.13\textwidth} p{0.03\textwidth} p{0.06\textwidth}@{}}
        \toprule
        Process Step & Failure Mode & Failure Effect & S & Failure Cause & O & Failure Measure & D & RPN \\
        \midrule
        Cell stacking & Incorrect cell placement & Misalignment of cooling system & 7 & Improper alignment guides & 5 & Adjust automated cell placement, visual inspection & 3 & 105 \\
        \midrule
        Welding of cell contact system & Weak weld joints & Increased resistance, potential overheating & 8 & Improper welding parameters, contaminated surfaces & 6 & Weld quality checks, resistance testing & 4 & 192 \\
        & Weld porosity & Reduced mechanical strength, potential failure & 9 & Contaminated surfaces, improper shielding gas flow & 5 & Weld X-ray inspection, surface cleaning procedures & 3 & 135 \\
        \bottomrule
    \end{tabular*}
\end{table*}

\begin{table*}[t]
    \centering
    \caption[Graph statistics]{Overview of $min$, $max$ and $average$ number of relationships for each label in the FMEA-KG. The KG has a total of 3,107 nodes and 4,576 relationships.}
    \label{tab:graph_data}
    \begin{tabular*}
    {\textwidth}{@{\extracolsep{\fill}}
    p{\dimexpr0.25\textwidth-2\tabcolsep\relax}
    p{\dimexpr0.25\textwidth-2\tabcolsep\relax}
    p{\dimexpr0.25\textwidth-2\tabcolsep\relax}
    p{\dimexpr0.25\textwidth-2\tabcolsep\relax}@{}}
        \toprule
        Label & $min$(\#relationships) & $max$(\#relationships) & $avg$(\#relationships) \\ \midrule
        \emph{FailureMode} & 2 & 11 & 3.45 \\
        \emph{FailureEffect} & 1 & 68 & 15.7 \\
        \emph{FailureCause} & 4 & 17 & 4.67 \\
        \emph{FailureMeasure} & 1 & 17 & 1.39 \\
        \emph{ProcessStep} & 2 & 71 & 20.30 \\ 
        \bottomrule
    \end{tabular*}
\end{table*}

This Section details the specifics of our implementation and the FMEA data. For our results, we chose an actual FMEA dataset from a production of high-voltage systems (HVS) from prismatic battery cells. Example data from the FMEA is presented in Table~\ref{tab:fmea_example_data}. With up to 35\% of the total production costs of an electric vehicle, HVS makes up the vehicle's most expensive component. Therefore, there is a great interest in the need for well-defined quality measurements \cite{kupper_future_2018}.

To construct the multi-label FMEA-KG with literals, we employed Neo4j \cite{neo4j_neo4j_nodate}, which provides us with its graph query language, Cypher.
The FMEA-KG comprises 3,107 nodes encompassing process steps, failure modes, failure effects, failure causes, and failure measures. A total of 4,576 relations connect the nodes. An overview of the number of relationships $min$, $max$, and $average$ for each label in the FMEA-KG can be found in Table~\ref{tab:graph_data}. The average number of relationships is between 1.39 (\emph{FailureMeasure}) and 20.30 (\emph{ProcessStep}). For example, there is a \emph{FailureMeasure} named "Wait for work instruction," which is related to 17 different \emph{FailureCauses}. 

The FMEA-KG contains 3,046 unique paths that can be traced from the root node \emph{FailureMode}. That means that for each \emph{FailureMode}, a path can be represented as a set of triples such that no two sets are identical. In other words, for each unique path $P$, there exists a set $S = \{ (n_1, r_1, n_2), (n_2, r_2, r_3), \ldots, (n_k, r_k, n_{k+1})\}$ where no other set $S'$ in the graph is identical to $S$. A sample path is presented in Table~\ref{tab:example_path}.

To compute the vector embedding of this path, we iterate over the connected nodes to retrieve their respective properties and concatenate them into a single string. For example:
\doublequotes{Process step: Cell stacking, Failure mode: Incorrect cell placement, Failure effect: Misalignment of the cooling system, S: 7 (...).}
The vector embedding is then computed using the text-embedding-ada-002 model \cite{greene_new_2022}.

Another important aspect is the instruction design for running the LLM's inference jobs. The instructions ensure that the LLM generates precise and relevant responses, enhancing the overall performance of context retrieval. The LLM of our KG-RAG framework is OpenAI's GPT-4 model with version 1106-Preview \cite{openai_gpt-4_2023}. We implement the KG-RAG framework's business logic as the backend service with a REST-API and employ a chat interface for the study (cf. Figure~\ref{fig:fmea_chatbot})\footnote{For those interested in reproducing the results with an example FMEA or exploring the framework further, the code for the RAG framework's backend service is available on GitHub at \url{https://github.com/lukasbahr/kg-rag-fmea}.}.

\begin{table*}[t]
\centering
\caption{Example path $P_1$ from the FMEA-KG illustrates a sequence of nodes and relationships that trace a failure mode, its effects, causes, and mitigation measures.}
 \label{tab:example_path}
\begin{tabular}{|p{12cm}|}
\hline
 $P_1 = $ 
 \{ \\ \hspace{0.5cm} (\textit{Cold Solder Joint}, \textit{occursAtProcessStep} \textit{Soldering Process}), \\ 
\hspace{0.5cm} (\textit{Cold Solder Joint}, \textit{resultsInFailureEffect}, \textit{Intermittent Connection}), \\ 
\hspace{0.5cm} (\textit{Cold Solder Joint}, \textit{isDueToFailureCause}, \textit{Inadequate Heating}),\\
\hspace{0.5cm} (\textit{Inadequate Heating}, \textit{isImprovedByFailureMeasure}, \textit{Increase Heating Temperature}) \\ \} \\
\hline
\end{tabular}
\end{table*}

\subsection{Study methodology} \label{subsec:methodology}
Although FMEA is a complex expert topic, information retrieval is relevant to all shop floor technical professionals involved in risk management.
In our study, we engaged 10 participants, including FMEA experts and non-experts, to evaluate the effectiveness of our KG-RAG framework.
The participants were assigned to extract information from the FMEA chatbot (cf. Section~\ref{subsec:implementation_details}) and the original FMEA-Excel spreadsheet as a baseline.
Assessing the performance of an information retrieval framework for FMEA presents unique challenges, as (i) there is no established baseline due to different FMEA approaches, (ii) the validity of the evaluation process in the study depends on the subject, and (iii) limited availability of domain experts. 
Nonetheless, in the context of FMEA, Excel as a comparative baseline is grounded in its widely used tool to work with FMEA in the manufacturing sector, particularly for small and medium-sized companies.

Throughout the study, each participant was asked to complete three distinct tasks, each designed to test the retrieval of different types of FMEA information. The tasks were designed to simulate scenarios professionals might encounter when dealing with FMEA on the shop floor. 
For example, \doublequotes{\emph{In step X process, there are faults due to Y. What prevention measure is provided to eliminate the fault?}}
For each task, the performance of the RAG framework and Excel was evaluated using five metrics: the correctness of the information retrieved, the usability of the interface, the relevance of the results to the task at hand, the completeness of the information provided, and the time taken to retrieve the necessary data. Employing these metrics allowed us, in addition to a qualitative assessment, to quantify and understand the advantages and potential drawbacks of the KG-RAG framework.

We give the subjects the following definitions for correctness, usability, relevance, completeness, and retrieval time.\\
\textbf{Correctness:} Evaluates the degree of precision of the information retrieved by the subject and whether it corresponds to the task requirement. For example, are the results based on information from the FMEA database or hallucinated by the model.\\
\textbf{Usability:} Measures the ease of use with which participants interact with the system interface. The interface is user-friendly, intuitive, and easy to navigate.\\
\textbf{Relevance:} Evaluates how closely the retrieved information aligns with the context of the question. Is there any unnecessary information for the task revealed?\\
\textbf{Completeness:} Indicates if the retrieved information provides a full and thorough answer to the task. For example, the result should match the inquiry for a task that asks for three failure causes.\\
\textbf{Retrieval time:} Measure the time and thus the efficiency of the retrieval process, particularly how long it takes a subject to obtain the information.

After each task, we asked the participants to assign a score (1-5) for each metric, reflecting their notion of how well they achieved the task with the FMEA chatbot and the Excel spreadsheet. We also asked them to share their thoughts and comments during and after each task and stopped the time.

\subsection{Results of the study} \label{subsec:results}

\begin{table*}[t]
    \centering
    \caption[Qualitative results of the study]{We asked $n=10$ participants to rate (1-5) each of the three tasks according to the depicted metrics for the FMEA chatbot and Excel. The results show the accumulated mean value with the standard deviation.}
    \label{tab:results_study}
    \begin{tabular*}
    {\textwidth}{@{\extracolsep{\fill}}
    p{\dimexpr0.150\textwidth-2\tabcolsep\relax}
    p{\dimexpr0.125\textwidth-2\tabcolsep\relax}
    p{\dimexpr0.125\textwidth-2\tabcolsep\relax}
    p{\dimexpr0.125\textwidth-2\tabcolsep\relax}
    p{\dimexpr0.125\textwidth-2\tabcolsep\relax}
    p{\dimexpr0.155\textwidth-2\tabcolsep\relax}
    p{\dimexpr0.195\textwidth-2\tabcolsep\relax}@{}}
        \toprule		
           & Correctness      & Usability       & Relevance       & Complete        & Retrieval Time  & Time Taken {[}min{]} \\
        \midrule
       Excel  & 4.38 $\pm$ 0.80 & 2.10 $\pm$ 0.70 & 3.95 $\pm$ 1.02 & 3.86 $\pm$ 1.10 & 2.76 $\pm$ 0.77 & 01:51 $\pm$ 01:26    \\
KG-RAG & 4.71 $\pm$ 0.71 & 4.71 $\pm$ 0.46 & 4.67 $\pm$ 0.91 & 4,38 $\pm$ 0.92 & 4.81 $\pm$ 0.40 & 01:19 $\pm$ 01:00    \\ 
        \bottomrule
    \end{tabular*}
\end{table*}

The results of the user experience design study have been positive throughout, and the FMEA chatbot was across the board higher evaluated, indicating an improvement over Excel to access FMEA information. The correctness of the information retrieved from the FMEA chatbot is evaluated at 7.53\% higher than Excel. Completeness and relevance achieve 13.47\% and 18.23\% higher results. Usability increased by 124.29\%. Although the measured time the participants took for each task is, on average, 00:32 minutes (24.72\%) shorter with the FMEA chatbot, their subjective perception of time differs with an evaluation of the retrieval time of 74.28\% higher compared to Excel. The highest standard deviation for the chatbot appears for completeness (0.92) and corresponds to the thoughts and comments we received from the participants. Table~\ref{tab:results_study} summarizes the results.

FMEA experts highlighted that the given context added to the answer helped to build trust in the chatbot's answer. The contexts provided sensible explanations, reduced the need for follow-up questions, and thus increased the relevance of the inquiry result.
\textit{\doublequotes{Context is great: gives a sensible explanation with more information. Saves to re-ask.}}
Other experts said that working with FMEA in Excel spreadsheets is not a pleasant task, for example, when seeking multiple explanations for a failure cause or gathering a general overview of the FMEA. 
One participant expressed: \textit{\doublequotes{The chatbot approach makes the FMEA much more interesting for acquiring knowledge and for simply using. So far, FMEA with Excel has not been used pleasantly, and you always need the help of the FMEA owners.}}

\begin{table}[t]
    \centering
    \caption[Results for the metrics for context retrieval]{Results on context recall (CR) and context precision (CP) for a baseline RAG implementation and our proposed KG-RAG framework with and without query search.}
    \label{tab:results_context_retrieval}
    \begin{tabular*}
    {0.48\textwidth}{@{\extracolsep{\fill}}
    p{\dimexpr0.27\textwidth-2\tabcolsep\relax}
    p{\dimexpr0.115\textwidth-2\tabcolsep\relax}
    p{\dimexpr0.15\textwidth-2\tabcolsep\relax}@{}}
    \toprule
     & CR & CP \\ 
     \midrule
    Baseline RAG (vector search) & 0.17 & 0.29 \\
    KG-RAG (vector search) & 0.22 & 0.44 \\
    KG-RAG  & 0.46 & 0.82 \\ 
    \bottomrule
    \end{tabular*}
\end{table}

All participants advertised the ease of use of the FMEA chatbot. The ten participants were already familiar with chat-based LLMs, which helped them adapt to the setting.
In particular, the ability to correctly interpret and respond to inquiries was highlighted. One user noted the proficiency of knowledge retrieval even though the input had vague language and, as a non-FMEA expert, did not use the precise terminology typically necessary for search queries in FMEA tools. \textit{\doublequotes{There was a spelling mistake in the query, which was relatively imprecise - I didn't use the precise term I was searching for but still got the correct answer.}} 
Another participant stated: \textit{\doublequotes{It's just cool that you don't have to be precise when describing and still get the right answer.}}

Although all participants had a positive impression of the proposed approach, some experts stated that a more thorough evaluation of the validity of the data is necessary before deploying it on the shop floor. Another remark from FMEA experts was that it would be beneficial to integrate the chatbot into an existing FMEA tool.

\subsection{Evaluation of the context retrieval} \label{subsec:online_evaluation}

\begin{table*}[t]
    \centering
    \caption[Example results with and without query]{Example results of the answers and retrieved contexts for the question, \textit{\doublequotes{How many failure effects with an S value of over 5 exist?}} and ground truth, \textit{\doublequotes{There are 14 failures that have a severity (S) score greater than 5.}} The context size is set to $k=3$. The question is formulated in particular to test the retrieval of numerical information.}
    \label{tab:example_data}
    \begin{tabular*}{\textwidth}{@{\extracolsep{\fill}}p{\dimexpr0.15\textwidth-2\tabcolsep\relax}p{\dimexpr0.28\textwidth-2\tabcolsep\relax}p{\dimexpr0.50\textwidth-2\tabcolsep\relax}@{}}
        \toprule		
         & Answer & Contexts \\
        \midrule
        Baseline RAG \newline (vector search) & There are a total of 3 failure effects with an S value of over 5. & [\doublequotes{3D Scanner 1 in deployment 3D System Generates pseudoIO related to defect images ...},  \doublequotes{Contact system CSK Lid mounting Issue Low voltage battery production ...}, \doublequotes{Contact system CSK Lid mounting Insulation issue 12 Particles ...}] \\
        \midrule
        KG-RAG \newline (vector search) & There is one failure with a S-value of over 5. & [\doublequotes{Process step: Form cell stacks, Failure effect: Scrap cells/module, S: 6 (…)}, \doublequotes{Process step: Form cell stacks, Failure effect: Process failure/increase throughput times, S: 3 (…)}, \doublequotes{Process step: Form cell stacks, Failure effect: no particular effects, S:1 (…)}] \\
        \midrule
        KG-RAG & There are 14 failures with a S score of over 5. & [\doublequotes{NumberFailureEffectsWithSOver5: 14}] \\
        \bottomrule
    \end{tabular*}
\end{table*}

The handling of FMEAs requires numerical data extraction to improve risk mitigation strategies. For example, experts may need to determine the primary cause of failure within a particular process step, imposing questions such as \textit{\doublequotes{What failure mode has the highest RPN for the process step X?}} In addition, experts also employ basic data analysis, such as identifying the process step with the mean RPN.
An FMEA Excel spreadsheet would entail a series of computational actions.
The merits of our proposed KG-RAG framework allow the retrieval of numerical data and basic statistical analysis of FMEA data.

To validate the effectiveness of our method, we evaluate the KG-RAG framework on a validation dataset comprising entries with questions and verified ground truth information about our FMEA data. These questions were a mix of open-ended questions (e.g., \textit{\doublequotes{How can we mitigate clocking in the cell stacking process?}}), yes/no questions (e.g., \textit{\doublequotes{Did the machine stop working due to overheating?}}), and ranking questions (e.g., \textit{\doublequotes{Rank the failure modes of the welding process step of the cell contact system by the risk priority number.}}).
We involved FMEA experts to ensure that the validation dataset is unbiased and that the ground truth data are accurately defined.
As a benchmark, we compare our KG-RAG framework against a baseline KG-RAG implementation that uses vector search only and a baseline RAG implementation with a random chunking strategy that operates on tabular structured data \cite{shuster_retrieval_2021}. 
The evaluation of the context retrieval is twofold, with the intent to verify both context recall and precision on the validation dataset.
Context recall (CR) measures the extent to which the retrieved context aligns with the ground truth data and is defined by
\begin{align}
\text{CR} = \frac{\text{\#ground truth attributable statements}}{\text{\#sentences in ground truth}},
\end{align}
for which \# is the number of total appearance \cite{shuster_retrieval_2021}.
Context precision (CP) is evaluated based on the ranking of the information within the context and its relevance. It is calculated by
\begin{align}
\begin{split}
\text{CP} &= \frac{1}{\text{\#relevant information}} \\ &\sum^n_{m=1} (\frac{\text{\#relevant information till $m$}}{m} \times r_m),
\end{split}
\end{align}
where $n$ denotes the total number of information in the context and $r_m$ is the binary value whether the information to the $m$-th piece of information is true ($r_m=1$) or false ($r_m=0$) \cite{salemi_evaluating_2024}.

Table~\ref{tab:results_context_retrieval} presents the results for CR and CP comparing the KG-RAG framework with and without query search and the baseline RAG implementation. The results show that with our approach, CR and CP (0.46, 0.82) are higher than without query search (0.22, 0.44) and the baseline RAG implementation (0.17, 0.29).
An examination of the retrieved context for the baseline RAG implementation and the KG-RAG framework with and without query search is shown in Table~\ref{tab:example_data}, for an example inquiry, \textit{\doublequotes{How many failure effects with a S value of over 5 exist?}} The context size is set to $k=3$. The contexts show that the method can correctly reason its answer from the incomplete context with vector search only. However, it could not correctly retrieve the number of failure effects from the FMEA data. The contexts of the baseline RAG are arbitrarily constructed, making it difficult for humans to trace the reasoning behind the answers.

\subsection{Discussion of the results} 
\label{subsec:discussion}

In developing a KG-enhanced RAG framework specifically tailored for FMEA, the results in Section~\ref{subsec:results} show significant advances in factual knowledge recall. The KG-RAG framework's improved ability to reason and comprehend vague queries enhances FMEA's accessibility and user-friendliness, benefiting experts and non-experts in extracting insights. In addition, reduced information retrieval time helps practitioners in the workplace by enabling agile decision-making and problem-solving.

The results in Section~\ref{subsec:online_evaluation} demonstrate that integrating a graph query search with a vector search enables basic analytical capabilities on datasets with mixed data types, such as FMEA, improving CR and CP. 
We further demonstrate that for datasets with a predefined structure, such as in expert domain data like FMEA, a well-defined chunking strategy can significantly enhance the results over random chunking strategies.
This and the context visualization within the chat interface foster confidence in the KG-RAG framework's capabilities. 
One of the main challenges we faced was dealing with abbreviations and specific terminology used in our FMEA. The FMEA contains many abbreviations and specific terminology unknown to the LLM, as some are undisclosed. To address this issue, we replaced less-known abbreviations (e.g., CSS) with their full forms (e.g., cell contacting system).

Although the study presents generally positive outcomes, the very nature of the domain fitting of our approach brings threats to validity. The challenge lies in obtaining statistical significance across metrics such as correctness, usability, relevance, completeness, and retrieval time, given the size of our user experience design study. In addition, conducting an automated evaluation proves challenging in the absence of the necessary expert knowledge, which complicates efforts to fully assess the performance of the KG-RAG framework.

On the technical level, further important aspects need to be explored. In particular, we did not test with alternative LLMs for text generation and vector embedding, as access to other online language models is restricted by data protection regulations. 
There are additional chunking strategies worth exploring. While we employ DFS to capture the underlying information of the FMEA-KG, other potentially more efficient path algorithms, such as breadth-first search or iterative deepening, are worth evaluating. Furthermore, other graph algorithms like community detection and centrality measures could provide deeper insights for summarization and analysis.

\subsubsection*{Discussion on handling epistemic uncertainty}
Uncertainty handling is a critical aspect of the KG-RAG framework, particularly in dealing with ambiguity and incomplete data. The framework employs a dual-layered validation process: graph query results are prioritized for their symbolic accuracy. At the same time, vector search outputs are supplemented with cosine distance metrics to signal semantic reliability. This layered approach ensures that uncertainty is clearly communicated to the user, reducing the risk of erroneous inferences.

The KG-RAG framework integrates human oversight as a key component to further mitigate risks. Users can independently validate AI outputs by presenting retrieved context and distance metrics. This mechanism fosters transparency and supports informed decision-making, particularly in complex, domain-specific scenarios like FMEAs in manufacturing.
The high usability and correctness score observed during the user experience design study underscores the effectiveness of these measures. Users expressed trust in the system due to its contextual clarity and the interpretability of uncertainty indicators. 

\subsubsection*{Broader implications of AI in risk management}
Integrating AI in risk management within industrial settings carries significant broader implications. As defined in ISO/IEC 31010 \cite{isoiec_risk_2019}, particular predictive quality has become prevalent in the manufacturing industry. Predictive quality aims to proactively assess risks and identify failures and defects before they occur, defining a new era of AI-enhanced risk management \cite{yazdi_navigating_2024}.
This, combined with recent developments in data-driven statistical modeling or integration of virtual reality and augmented reality technologies, provides practitioners greater insights into the production process, improved predictive capabilities, and proactive risk assessment \cite{shang_data_2019, yazdi_augmented_2024}.

Using chatbots for risk management in manufacturing carries important implications that must be addressed. Before implementing chatbots, it is essential to examine ethical considerations, data privacy, and the importance of critical oversight.
First and foremost, user safety must be ensured. For FMEAs that guide failure modes with potentially dangerous remedial measures, verifying information with domain experts is crucial to avoid any risk or harm to users.
Second, user privacy must be protected, as interactions with chatbots may involve sensitive information.
Lastly, every decision made by the chatbot requires trained users, particularly on the shop floor, and continuous human oversight to ensure safe and accurate operation.
\section{Conclusion}
\label{sec:conclusion}
In this paper, we propose a KG-RAG framework that synergies KG with LLM to enhance the QA capabilities within the FMEA domain. 
For that, we formalize a set-theoretic standardization of the FMEA, allowing any FMEA to be transposed into KG. Building on this standardization, we develop a schema to model the FMEA data as a multi-label KG with literals. To compute the vector embeddings of the FMEA, we suggest an algorithm to fully capture the information from the KG. We then offer a solution for a novel KG-RAG framework that can retrieve FMEA information and allow for analytical inquiries. 

The results of the user experience design study suggest a promising direction for the retrieval and interpretation of the semantic content of the FMEA data, enabling a more nuanced and analytical approach to risk assessment. Additionally, the evaluation of the context retrieval demonstrates that retrieval of numerical data is improved when augmented with the graph query interface of the KG.
The findings indicate that a general integration of LLMs into the shop floor could advance quality strategies.

However, in the future, many questions remain open. For one, the work's focus is on the retrieval of information, thereby leaving the open investigation to other LLMs for their effectiveness in generating and computing vector embeddings. 
A suite of graph algorithms could further refine chunking strategies for KG-built RAG applications.
Moreover, the framework can be extended to support guided user input, which could significantly improve the data quality of the FMEA database. 
Other thinkable investigations are around expanding the knowledge base to encompass other quality management data, such as PDCA or manufacturing data, which could allow for a broader application of a KG-RAG framework across different quality assurance contexts.

\section*{Funding}
This work was supported by the Bavarian Ministry of Economic Affairs, Regional Development, and Energy, the BayVFP's \doublequotes{Digitalization} funding line, the KIProQua project, and the BMW Group.

\section*{CRediT authorship contribution statement}
\textbf{Lukas Bahr}: Writing – original draft, Visualization, Validation, Software, Methodology, Investigation, Data curation, Conceptualization.
\textbf{Christoph Wehner}: Writing – original draft, Validation, Methodology, Conceptualization.
\textbf{Judith Wewerka}: Writing – review \& editing, Validation, Investigation.
\textbf{José Bittencourt}: Validation, Funding acquisition, Data curation.
\textbf{Ute Schmid}: Funding acquisition, Conceptualization
\textbf{Rüdiger Daub}: Writing – review \& editing, Supervision, Project
administration, Conceptualization.

\section*{Declaration of competing interest}
The authors declare the following financial interests/personal relationships which may be considered as potential competing interests:
Lukas Bahr reports financial support, administrative support, article publishing charges, and equipment, drugs, or supplies were provided by BMW Group. Christoph Wehner reports financial support was provided by the Bavarian Ministry of Economic Affairs, Regional Development, and Energy. José Bittencourt and Judith Wewerka are employed at BMW Group. If there are other authors, they declare that they have no known competing financial interests or personal relationships that could have appeared to influence the work reported in this paper.

\section*{Acknowledgements} \label{sec:acknowledgements}
The authors would like to thank the engineers and co-workers on the shop floor who participated in the evaluation of the KG-RAG framework. The authors further gratefully acknowledge the financial support from the Bavarian Ministry of Economic Affairs and the BMW Group.

\section*{Data availability}
The authors do not have permission to share data.

\bibliographystyle{cas-model2-names}

\bibliography{bib}

\end{document}